\documentclass[journal]{IEEEtran} 

\usepackage{cite}
\usepackage{graphicx}
\usepackage{subfig}
\usepackage{listings}

\begin{document}
\bibliographystyle{plain}

\title{A CAD Interface for GEANT4}

\author{Christopher~M~Poole, Iwan~Cornelius, Jamie~V~Trapp, Christian~M~Langton%
\thanks{Correspondence can be directed to \texttt{christopher.poole@qut.edu.au}}
\thanks{All authors are with the Discipline of Physics, Faculty of Science and Technology, and the Institute of Health and Biomedical Innovation at Queensland University of Technology, Brisbane, Australia.}
\thanks{This work is funded by the Queensland Cancer Physics Collaborative, and Cancer Australia (Department of Health and Ageing) Research Grant 614217.}}

\maketitle

\begin{abstract}
Typically used as a tool for Monte Carlo simulation of high energy physics experiments, GEANT4 is increasingly being employed for the simulation of complex radiotherapy treatments.
Often the specification of components within a clinical linear accelerator treatment head is provided in a CAD file format.
Direct import of these CAD files into GEANT4 may not be possible, and complex components such as individual leaves within a multi-leaf collimator may be difficult to define via other means.
Solutions that allow for users to work around the limited support in the GEANT4 toolkit for loading predefined CAD geometries has been presented by others, however these solutions require intermediate file format conversion using commercial software.
Here within we describe a technique that allows for CAD models to be directly loaded as geometry without the need for commercial software and intermediate file format conversion.
Robustness of the interface was tested using a set of CAD models of various complexity;  for the models used in testing, no import errors were reported and all geometry was found to be navigable by GEANT4.
\end{abstract}

\section{Introduction}
\IEEEPARstart{G}{eometry} and Tracking (GEANT4) is a C++ toolkit specifically designed to track particles traversing a geometry whilst being subject to physical processes, it finds application in fields such as nuclear and particle physics and space engineering, with increasing use in medical physics \cite{agostinelli2003geant4,caccia2010medlinac2, jan2011gate, spezi2008overview, grevillot2011simulation}.
Numerous physical processes can be modeled including photo-nuclear interactions, optical processes such as scintillation and Cherenkov radiation and other particle interactions over a wide energy range ($250\ eV$ up to $TeV$ energies); the full gamut of processes available to the user is described by others \cite{agostinelli2003geant4}.
Additionally, the toolkit provides functionality for the inclusion and exclusion of the desired processes as well as sufficient extensibility to included custom or user defined processes.
Fast and effective geometry definition is also available to the user for relatively simple objects using constructs such as \texttt{G4Orb} for defining orbs or spheres, \texttt{G4Box} for defining rectangular prisms and the concept of boolean solids \cite{agostinelli2003geant4}.
At the most fundamental level, the GEANT4 geometry hierarchy is divided into solids, logical volumes and physical volumes where solids described shape, logical volumes define material properties and mother daughter relations, and physical volumes define placement within the mother volume.

Geometry Description Mark-up Language (GDML) is a comprehensive geometry description format using Extensible Mark-up Language (XML) that allows for the persistence of many aspects of a geometry, including material properties and assemblies \cite{chytracek2006geometry}.
Functionality within the toolkit readily allows for geometries to be saved and reloaded as GDML, however there is little support for converting pre-existing user defined Computer Aided Design (CAD) models to GDML or indeed directly loading these same CAD models as geometry \cite{constantin2010linking}.
Of the two methods available to the user for accomplishing this task, both are reliant of intermediate file format conversion using commercial software \cite{constantin2010linking,beutier2003fastrad}.

\begin{figure*} 
  \centering
  \subfloat[][]{\label{fig:flow1}\includegraphics[]{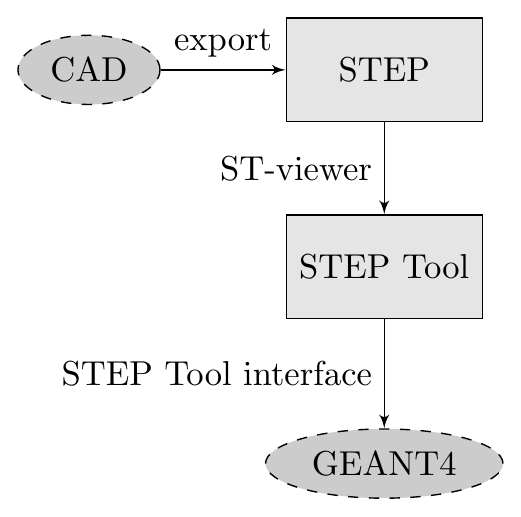}}                
  \subfloat[][]{\label{fig:flow2}\includegraphics[]{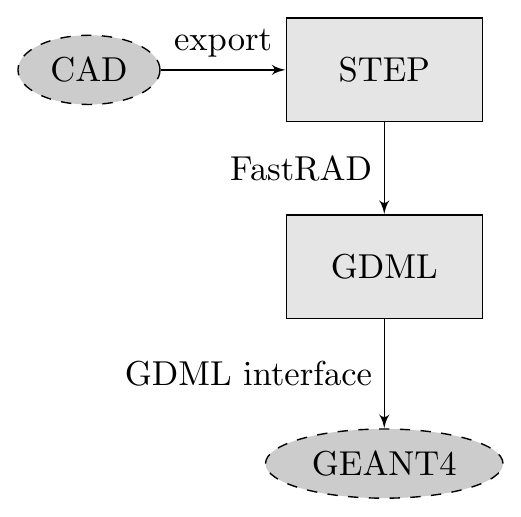}}
  \subfloat[][]{\label{fig:flow3}\includegraphics[]{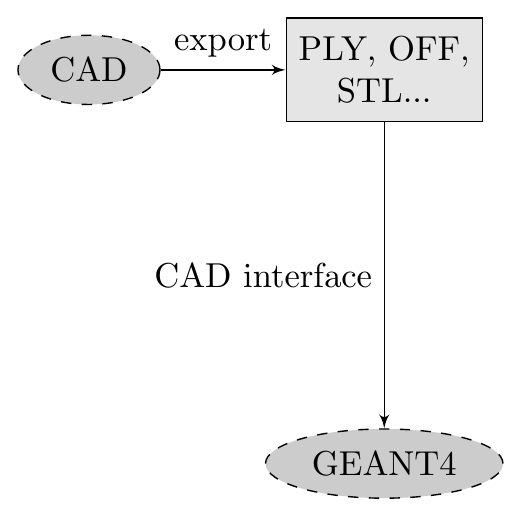}}
  \caption{A diagrammatic comparison between two currently available CAD import techniques where \subref{fig:flow1} shows the technique described in persistence example G02, \subref{fig:flow2} shows the STEP to GDML conversion technique \cite{constantin2010linking}, and \subref{fig:flow3} shows the new direct import technique proposed here within.}
  \label{fig:flowmain}
\end{figure*}

Many CAD packages export files compliant to the standard for the exchange of product model data (STEP - ISO 10303), a standard designed to supersede the still widely used initial graphics exchange specification (IGES) \cite{pratt2001introduction, iges2007}.
As such, ISO standard STEP has been the target format for loading CAD geometry into GEANT4.
Persistence example, with identifier `G02', distributed with GEANT4 describes loading STEP Tools (STEP Tools Incorporated, New York) files directly, with the intermediate conversion of STEP to STEP Tools using the commercial ST-Viewer program or ST-Developer libraries (STEP Tools Incorporated, New York), refer to \figurename \ref{fig:flowmain}\subref{fig:flow1}.
This process allows for assemblies of components to be loaded directly, however the STEP Tools programs and libraries may be prohibitively expensive for some users.
Constantine et al described the process of converting STEP to GDML using the commercial software FastRad (Tests \& Radiation - Toulouse)\cite{beutier2003fastrad}, refer to \figurename \ref{fig:flowmain}\subref{fig:flow2}.
FastRad again may be considered prohibitively expensive for some users with the requirement of annual licensing, in addition to this, the trial version limits the conversion process to no more than 20 elements per assembly.

Principally, CAD file formats such as stereo-lithography format (STL) or Stanford polygon file format (PLY) describing a 3-dimensional volume use tessellated polygon meshes (usually triangular or quadrangular) to define its closed surface.
A cloud of points in 3-dimensional space define the vertexes of the mesh and a collection of faces define the interconnects between the vertexes.
The equivalent geometric construct in GEANT4 to this scheme is the tessellated solid (\texttt{G4TessellatedSolid}); except for the two previously mentioned techniques, mappings between the \texttt{G4TessellatedSolid} construct and CAD file formats are not readily available\cite{constantin2010linking}.
Programming toolkits however, already exist for the creation and manipulation of triangular and quadrangular meshes; the kind exportable by many modern CAD packages \cite{amenta2001power, vcglib2011online}.
Most notably the templated C++ library VCGLIB (Visual Computing Laboratory, Italy) offers advanced mesh manipulation routines and point cloud surface reconstruction algorithms.
In addition to this, VCGLIB offers an import/export interface for many common CAD file formats \cite{vcglib2011online}.

Here within we describe a very simple technique that allows for CAD models to be directly loaded as geometry in GEANT4 using VCGLIB.
This technique is capable of handling many file formates exportable by CAD programs whilst not relying on intermediate file format conversions.

\section{Methods}
    \subsection{GEANT4 CAD Interface}
    \label{sec:cad_interface}
    Three C++ classes were created describing a vertex, face and a triangular mesh.
    The class \texttt{CADVertex} of base type \texttt{vcg::Vertex}  was used to store a single vertex defined by three double precision values $(x,\ y,\ z)$.
    For faces, the class \texttt{CADFace} of base type \texttt{vcg::Face} was used to store a reference to all vertices that define the face $(a,\ b,\ c)$.
    Finally, the class \texttt{CADMesh} templated on two \texttt{std::vector} containers was created.
    The first vector container was templated on the custom type \texttt{CADVertex}, and the second on the custom type \texttt{CADFace} - in effect the \texttt{CADMesh} type contained two collections, one describing all the vertices in a mesh and another describing all of the interconnect between the vertexes (the faces). 
    Through an instance of a \texttt{CADMesh}, the description of a triangular mesh with a cloud of vertices interconnected by a group of faces could be created.
    
    Wrappers in VCGLIB extend the VCGLIB core to provide additional mesh manipulation functionality including importers and exporters.
    Using the \texttt{Open} method in the wrapper \texttt{vcg::tri::io::ImporterPLY} and parsing to it a PLY file and a reference to an instance of \texttt{CADMesh}, the instance of \texttt{CADMesh} was populated with a 3-dimensional triangular facet mesh described the PLY file.
    Other triangular facet mesh importers include OBJ, OFF, and STL file formats \cite{vcglib2011online}.
    Random access to faces in the mesh was available using the \texttt{CADMesh::FaceIterator} iterator where each face provided a pointer to each vertex $(a,\ b,\ c)$ of the current face and each vertex provided pointers to  its coordinates $(x,\ y,\ z)$.
    
    For each \texttt{CADMesh} a \texttt{G4TessellatedSolid} was initialised in GEANT4 (version 9.4, patch 01, $25^{th}$ February 2011).
    Using the \texttt{CADMesh::FaceIterator} iterator, the \texttt{AddFacet} method of the \texttt{G4TessellatedSolid} was called once for each mesh face.
    The \texttt{AddFacet} method took three \texttt{G4ThreeVector} arguments describing the coordinates of each vertex that define the current face.
    Once the iterator looped over all faces, the \texttt{G4TessellatedSolid::SetSolidClosed} method was called, preventing further faces from being added to the volume.
    The \texttt{G4TessellatedSolid}, was now available for material assignment and placement within the GEANT4 user geometry using standard programmatic techniques in GEANT4.
    
    At the user level, all functionality provided by the CAD interface could be incorporated into any pre-existing geometry with the inclusion of a single header file in the user detector construction.
    Initialisation of a \texttt{CADMesh} object along with a volume described in a CAD file, would provide a \texttt{G4TessellatedSolid} suitable for inclusion in the user geometry in the same manner as with any typical \texttt{G4Solid} object.
    Following the standard GEANT4 geometry hierarchy, material properties and volume meta-data unique to the volume could be assigned with association to a \texttt{G4LogicalVolume}, and placement of the \texttt{G4TessellatedSolid} within a mother volume with association to a \texttt{G4PhysicalVolume}, see code listing \ref{code:example}.
    
    \lstset{language=C++, basicstyle=\footnotesize, xleftmargin=20pt, xrightmargin=20pt, numbers=left, stepnumber=1 , frame=trBL}
    \begin{lstlisting}[float, caption=Basic usage of the CADMesh class in a user detector constructor, label=code:example]
#include "CADMesh.hh"
...
CADMesh mesh;
G4VSolid * solid;
G4LogicalVolume * logical;
G4VPhysicalVolume * physical;
...
solid = mesh.LoadMesh("sphere.stl", "STL");
logical = new G4LogicalVolume(solid, water,
          "logical", 0, 0, 0);
physical = new G4PVPlacement(0,
           G4ThreeVector(), logical,
           "physical", world, false, 0);
    \end{lstlisting}
    
    \subsection{Test Cases}
    Six test volumes were used  to verify the performance of the GEANT4 CAD interface.
    Three simple geometries, a truncated cone and a sphere generated using MeshLab (Visual Computing Laboratory, Italy) and an artificial hip (generated in-house) were saved in all formats capable of import by the CAD interface.
    Additionally, a flattening filter from a Varian clinical linear accelerator, a single leaf from a Varian multi-leaf collimator and a model of a pelvis from a CIRS Pelvic Phantom (Model 048) were loaded into GEANT4 using the CAD interface \cite{poole2011vectorised}.
    Each volume, when loaded was visually inspected for qualitative geometric integrity using the GEANT4 OpenGL viewer.
    Tessellated solid meta-data was dumped for each solid using the  \texttt{G4TessellatedSolid::DumpInfo} method providing a list of each face and vertex coordinate used to define the GEANT4 volume.
    This data was then compared directly to the original CAD file describing the same volume.
    A pass level of 100\% matching was set, no rounding errors of the vertex coordinates were accepted and vertex and face counts had to agree exactly.
    
    So as to ensure the loaded geometries were navigable by the GEANT4 kernel, each volume was assigned the material \texttt{G4\_WATER} and positioned at the center of a \texttt{G4\_AIR} filled world volume.
    A general particle source (GPS) was initialised and configured to produce a beam of geantinos (the standard GEANT4 debugging pseudo-particle) aimed at the test volume.
    Each test volume was bombarded with $100,000$ geantinos with the tracking verbosity level set to one and the step length set to $0.1\ mm$, ensuring any navigation errors associated with the imported CAD geometry were output; the angular distribution of the beam was set so as to target the entire volume ensuring all faces were inside the beam.
    At any time, if the navigator was unable to determine if it was inside or outside of the tessellated volume as a consequence of invalid volume definition, the test was failed.
    
\section{Results}
\begin{figure*}
  \centering
  \subfloat[cone]{\label{fig:cone}\includegraphics[width=100.0px]{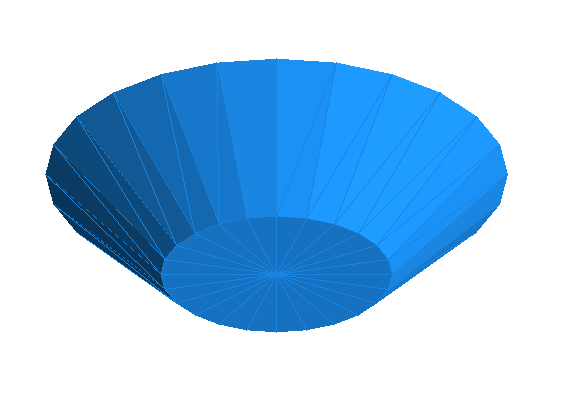}}                
  \subfloat[sphere]{\label{fig:sphere}\includegraphics[width=100.0px]{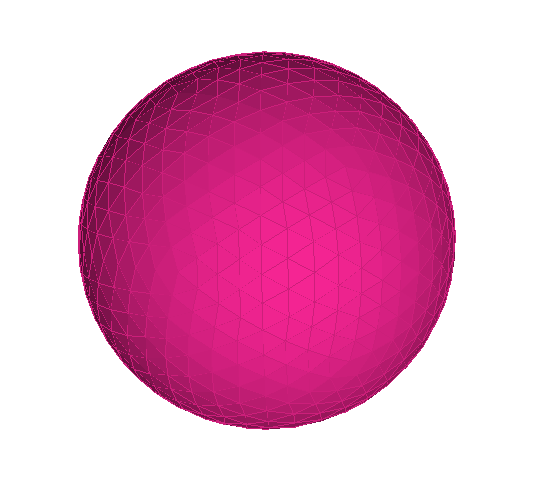}}
  \subfloat[hip prosthesis]{\label{fig:hip}\includegraphics[width=100.0px]{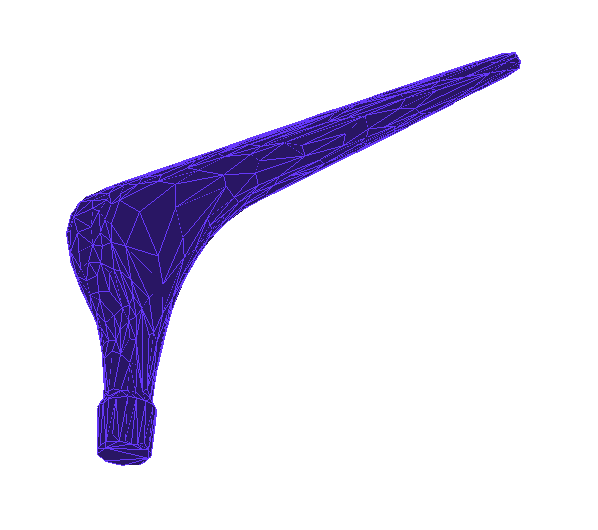}}
  
  \subfloat[flattening filter]{\label{fig:flatfilt}\includegraphics[width=100.0px]{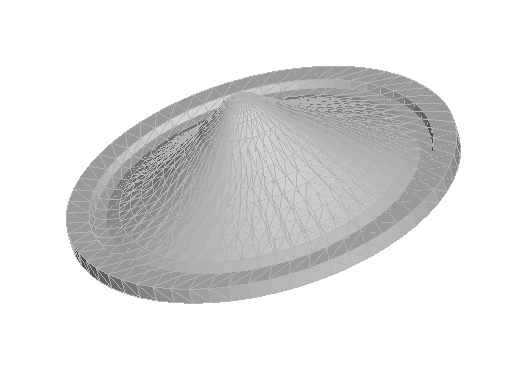}}                
  \subfloat[MLC leaf]{\label{fig:mlc}\includegraphics[width=100.0px]{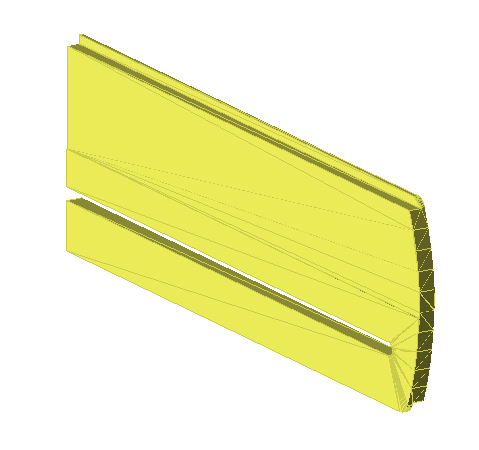}}
  \subfloat[pelvis model]{\label{fig:pelvis}\includegraphics[width=100.0px]{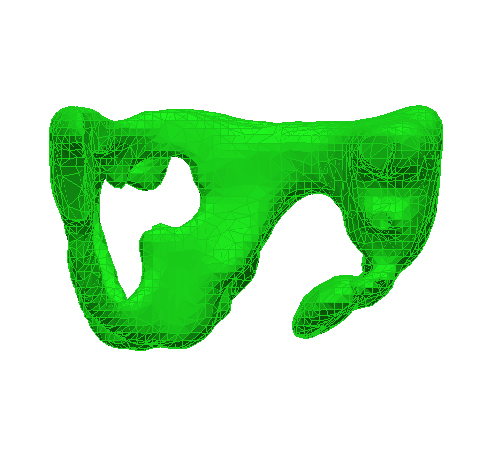}}
  \caption{Six test geometries loaded directly into GEANT4 using the proposed CAD interface.}
  \label{fig:geometries}
\end{figure*}

\begin{table}
\centering
\begin{tabular}{lrrcc}
Name 	& Verts & Faces & Inspection & Dump/Navigation\\
\hline\hline
cone	& 50	& 96 	& Pass & Pass	\\
sphere	& 642	& 1280	& Pass & Pass	\\
hip	& 502	& 1000	& Pass & Pass	\\
leaf	& 120	& 236 & Pass & Pass	\\
filter	& 1235	& 2346	& Pass & Pass	\\
pelvis	& 4986	& 10000 & Pass & Pass	\\
\hline
\end{tabular}
\caption{Properties of the six volume loaded into GEANT4 via the proposed CAD interface.}
\label{tab:quality} 
\end{table}

\figurename \ref{fig:geometries}\subref{fig:cone}, \figurename \ref{fig:geometries}\subref{fig:sphere} and  \figurename\ref{fig:geometries}\subref{fig:hip} show three simple geometries loaded into GEANT4 via the CAD interface described in section \ref{sec:cad_interface}; the geometries were visualised using the GEANT4 OpenGL viewer.
Further to this, \figurename \ref{fig:geometries}\subref{fig:flatfilt}, \figurename \ref{fig:geometries}\subref{fig:mlc} and \figurename\ref{fig:geometries}\subref{fig:pelvis} show more complex geometries loaded into GEANT4.
All test volumes imported using the CAD interface, independent of the CAD file format, passed qualitative visual inspection - no corruption of the geometry was visible.
Table \ref{tab:quality} displays the vertex and face counts for each volume; there was no difference between the vertex and facet counts reported by VCGLIB or GEANT4.
A comparison between the original CAD file and the dumped face and vertex information from each tessellated solid generated no errors.
When bombarded with a simulated source of geantinos, no navigation errors were reported.
In the case that a face were to be purposefully excluded from the \texttt{G4TessellatedSolid}, the volume would be unnavigable as a consequence of an undefined `inside' or `outside'; a non-closed solid.

\section{Discussion \& Conclusion}
Using the templated C++ mesh manipulation library VCGLIB, we have demonstrated a technique whereby CAD models may be directly imported as geometry into GEANT4 without the express need for file format conversion using commercial software.
\figurename \ref{fig:geometries} shows a collection of CAD models successfully loaded into GEANT4 via the CAD interface.
Reliability of the interface in terms of preserving model integrity during import was evaluated quantitatively by comparing the vertex coordinates as reported by GEANT4 to the actual vertex coordinates described in the source CAD file.
Whilst the interface does not parse material properties or other GEANT4 specific meta-data, a CAD model imported using the interface may be saved as part of an assembly in GDML using tools that are already part of the toolkit.
By removing the intermediate file format conversion step, unnecessary expenditure on commercial third party software can be avoided and any CAD model described by a triangular tessellated surface may be directly loaded as geometry within GEANT4.
Further work will focus on increasing the number of CAD file formats available for direct import using the technique as well as enabling support for assemblies and the preservation of volume specific meta-data such as material composition.
Currently this work is part of software tool using GEANT4 for the simulation of clinical linear accelerators \cite{cornelius2011commissioning}, which is freely available (along with accompanying documentation) upon request made to the corresponding author.
Source code for the CAD interface described here within may be acquired from: \texttt{https://code.google.com/p/cadmesh/}

\IEEEtriggeratref{11} 
\nocite{*}
\bibliography{main}

\end{document}